\documentclass[12pt,draftcls,onecolumn]{IEEEtran}
\usepackage[utf8]{inputenc}
\usepackage{amsmath,mathtools}
\usepackage{amsthm}
\usepackage{amsfonts}
\usepackage{amssymb}
\usepackage{cite} % For collocating the continuous reference
\usepackage{array} % Withoit this the \newcolumntype was not working
\usepackage{easyReview} % for \comment, \alert, etc.
\usepackage{soulutf8} % for accented letters
\newtheorem{theorem}{Theorem}[]
\newtheorem{lemma}[]{Lemma}
\usepackage{algorithm,algorithmic}
\usepackage{caption}
\usepackage{url}
\usepackage[list=true]{subcaption}
\newcolumntype{C}[1]{>{\centering\let\newline\\\arraybackslash\hspace{0pt}}m{#1}}
\DeclarePairedDelimiter\abs{\lvert}{\rvert}
\DeclarePairedDelimiter\norm{\lVert}{\rVert}
\DeclarePairedDelimiter\curleybracket{\lbrace}{\rbrace}
\begin{document}
	\bstctlcite{IEEEexample:BSTcontrol}
	\title{Instantaneous Channel Oblivious Phase Shift Design for an IRS-Assisted SIMO System with Quantized Phase Shift}
	\author{Shashank Shekhar$^{1}$, Athira Subhash$^{1}$, Tejesh Kella$^{2}$, and Sheetal Kalyani$^{1}$
		\thanks{
			1. Shashank Shekhar, Athira Subhash, and Sheetal Kalyani are with the Dept. of Electrical Engineering, Indian Institute of Technology, Madras, India. Emails: \{ee17d022@smail,ee16d027@smail, and skalyani@ee\} .iitm.ac.in. 
			\\ 2. Tejesh Kella is with Qualcomm India Private Ltd.,
			Hyderabad 500081, India (email: kellatejesh@gmail.com)}
	}
	\maketitle 
	
	\begin{abstract}
		We design the phase shifts of an intelligent reflecting surface (IRS)-assisted single-input-multiple-output communication system to minimize the outage probability (OP) and to maximize the ergodic rate. Our phase shifts design uses only statistical channel state information since these depend only on the large-scale fading coefficients; the obtained phase shift design remains valid for a longer time frame. We further assume that one has access to only quantized phase values. The closed-form expressions for OP and ergodic rate are derived for the considered system. Next, two optimization problems are formulated to choose the phase shifts of IRS such that (i) OP is minimized and (ii) the ergodic rate is maximized. We used the multi-valued particle swarm optimization (MPSO) and particle swarm optimization (PSO) algorithms to solve the optimization problems. Numerical simulations are performed to study the impact of various parameters on the OP and ergodic rate. We also discuss signaling overhead between BS and IRS controller. It is shown that the overhead can be reduced up to $99.69 \%$ by using statistical CSI for phase shift design and $5$ bits to represent the phase shifts without significantly compromising on the performance.
	\end{abstract}
	\section{Introduction}
	Intelligence reflecting surface (IRS) has immense potential to enhance the performance of existing wireless communication systems by introducing desired phase shifts to the incident wave \cite{liaskos2018new}. Metamaterials can change the properties of incident electromagnetic (EM) waves in a programmed manner. A hypersurface that consists of several independent meta-atoms is called IRS. These meta-atoms, also known as IRS elements, can configure the incident EM wave by changing its amplitude and phase \cite{wu2019towards}. It is envisioned that the IRS will be an essential facilitator for future generations of wireless communication \cite{bariah2020prospective,saad2019vision,akyildiz20206g}. Several works studied the IRS with the collaboration of many other technologies, such as index modulation \cite{gopi2020intelligent}, non-orthogonal multiple access (NOMA) \cite{cheng2021downlink}, and full-duplex (FD) \cite{zhang2020sum}, to name a few. Works like \cite{tao2020performance,de2021large,atapattu2020reconfigurable,kudathanthirige_icc_20,charishma2021outage} have focused on characterizing outage probability (OP) for IRS-assisted single-input-single-output (SISO) systems. Recently, authors in \cite{jayalal2022sinr} characterized the SINR of an IRS-assisted multi-user multiple-input-single-output (MISO) system.
	\par The authors in \cite{wu2019intelligent} used alternating optimization to jointly optimize the phase shift at IRS and active beamforming at the base station (BS) to minimize the transmit power at BS while ensuring a minimum signal-to-interference-plus-noise ratio (SINR) threshold at each user. Later in \cite{wu2020beamforming}, they extended their work to the scenario where only finite phase shifts are available to the IRS. The downlink of a single-user IRS-assisted MISO system was considered in \cite{mishra2019channel}, where a closed-form near-optimal phase shift design has been proposed based on instantaneous CSI and continuous phase shift at the IRS.  In \cite{yu2019miso}, the spectral efficiency of the IRS-assisted MISO system was maximized by jointly designing the beamformer at BS and phase shift at IRS. Joint optimization of beamforming at BS and phase shift at IRS was performed to minimize the user's instantaneous OP in \cite{fang2020outage}. The downlink of single user IRS-assisted MISO system was considered in \cite{feng2020deep}, and the phase shift design was proposed to maximize instantaneous received SNR using the deep reinforcement learning framework. In \cite{abeywickrama2020intelligent}, authors proposed a relation between phase shift and reflection coefficient at IRS. Next, considering a similar system model as in \cite{feng2020deep} with multiple users, they used a penalty-based algorithm to solve for transmit beamforming and phase shift at IRS such that the transmit power is minimum.
	\par In \cite{guo2020outage}, the authors have analyzed and minimized the OP of the IRS-assisted MISO system with a deterministic BS-IRS link. 
	A closed-form expression for optimal beamforming vector is derived in \cite{subhash2022max} for  the IRS-assisted SIMO system with multiple users. They alternatively used geometric programming and the matrix-lifting method for power allocation and phase shift design at IRS such that the minimum SINR is maximized, respectively. In \cite{subhash2022optimal}, authors worked on a similar system model as in \cite{subhash2022max}, and they derived an equation for solving the asymptotic minimum SINR using the tools from random matrix theory. They employed alternating optimization to solve for the beamforming vectors at BS, power allocation of users, and phase shift at IRS to maximize the minimum SINR, considering the availability of continuous phase shift at IRS.
	In \cite{dai2021statistical,zhi2021statistical}, authors studied the IRS-assisted multi-user multiple-input-multiple-output (MIMO) system with and without hardware impairments, respectively. They focused on ergodic rate and derived an approximation based on statistical CSI. Then, the phase shift design is proposed using genetic algorithms without considering the impact of quantized phase shifts at the IRS. In \cite{han2019large}, authors derived a statistical upper-bound (UB) on the ergodic rate using Jenson's inequality for an IRS-assisted MISO system and proposed a phase shift design that maximizes the derived UB on the ergodic rate. Mathematically, it is equivalent to maximizing the mean of SNR.  
	
	\par It is clear that proper phase shift design at IRS is extremely important for the      effective use of the IRS. However, a phase shift design strategy based on instantaneous CSI comes with a feedback overhead since it requires updating the phase shift design for each small-scale fading coherence interval and giving feedback to IRS. Most of the previous works, such as \cite{wu2019intelligent,yu2019miso,fang2020outage,feng2020deep,abeywickrama2020intelligent,wu2020beamforming,subhash2022max} focus on phase shift design with the knowledge of instantaneous channel state information (CSI). Instead, the statistical parameters of the channel link depend on the large-scale fading coefficients, which vary slowly over time and may remain the same for at least $ 40 $ small-scale fading coherence intervals \cite{ashikhmin2018interference,rappaport1996wireless, Ngo2017:CellFree}. The IRS elements are programmed or controlled with the help of an IRS controller, and the BS communicates with the IRS controller over a separate wireless link to program the phase shift of the IRS elements \cite{wu2019towards,abeywickrama2020intelligent}. The statistical parameter-based phase shift design approach  will require lesser feedback between BS and IRS controller when compared with an instantaneous CSI-based scheme. The works like \cite{guo2020outage,subhash2022optimal,dai2021statistical,zhi2021statistical} consider the statistical CSI for designing the phase shift at the IRS with the assumption that any continuous phase value can be assigned to IRS elements. However, the IRS is envisioned to be a low-cost passive device with a high number of reflective elements; hence the availability of infinite resolution phase shift at elements of the IRS is not practical due to hardware limitations \cite{wu2019towards,han2019large,wu2020beamforming}, there can be only a finite number of discrete values among which the IRS has to select the phase shift. Thus, it is essential to study the impact of available quantization levels at the IRS. Hence, for the practical implementation of the IRS, we need to focus on two things. 1) design the phase shift using statistical CSI so that the feedback overhead cost is low between BS and IRS controller, and 2) consider only the availability of finite phase shift level at IRS elements to ensure the low cost of the IRS device.
	\par Motivated by the above reasons, in this work, we have considered an uplink IRS-assisted SIMO system where the phase shift design is done using the statistical CSI, and the effect of quantized phase shift is also studied. Our focus is to minimize the OP, which requires the characterization of end-to-end SNR, unlike ergodic rate in \cite{han2019large}, which can be upper bounded by just knowledge of the mean of SNR. The main contribution of this work can be summarized as follows
	\begin{itemize}
		\item We approximated the end-to-end SNR  of the uplink of the IRS-assisted single-user SIMO system by a Gamma RV using the moment matching technique.
		\item We derive the closed-form approximation for OP and ergodic rate for the considered system. Using these closed-form expressions, we formulated the optimization problem for the minimization of OP and the maximization of the rate. Next, we solved the formulated optimization problem using multi-valued discrete particle swarm optimization (MPSO) and particle swarm optimization (PSO) algorithms.  
		\item The phase shift design obtained through the formulated optimization problems depends on the large-scale fading coefficients. Hence, it requires less-frequent reconfiguration of the IRS, and that reduces the feedback overhead. 
		\item We study the impact of various system parameters, such as the number of elements at the IRS, the number of bits available for quantization at the IRS, the number of antennas at the BS, and the transmitted power.
		\item We show that a significant reduction in overhead is achieved with our scheme as discussed in Section \ref{Sec: MISO_IRS_Rician_Overhead}.
	\end{itemize}
	\begin{table}[!t]
		\centering
		\begin{tabular}{|C{1.5cm}|C{1.5cm}|C{2.5cm}|C{1.5cm}|C{2.5cm}|C{3cm}|}
			\hline 
			Reference & Antenna Model & Quantized Phase Shifts & Statistical CSI & Objective & Optimization Methodology \\
			\hline 
			\cite{wu2019intelligent} &  MISO  &  \texttimes  & \texttimes &  Transmit power minimization  & semidefinite relaxation (SDR) and alternative optimization  \\ \hline
			\cite{wu2020beamforming} & MISO   & $\checkmark$  & \texttimes  & Transmit power minimization  & Successive refinement algorithm  \\ \hline
			\cite{mishra2019channel} & MISO &  \texttimes &  \texttimes & received SNR maximization & Analytical \\ \hline 
			\cite{yu2019miso} &  MISO  & \texttimes  & \texttimes  & SE maximization  &  Fixed point iteration and manifold optimization \\ \hline
			\cite{fang2020outage} &  MISO  & \texttimes & \texttimes & OP minimization  & Stochastic gradient descent (SGD)  \\ 
			\hline 
			\cite{feng2020deep} &  MISO & \texttimes & \texttimes & received SNR maximization & Deep reinforcement learning \\ \hline 
			\cite{abeywickrama2020intelligent} & MISO  & \texttimes & \texttimes   &  Transmit power minimization & penalty-based algorithm \\ \hline
			\cite{guo2020outage  }&  MISO  & \texttimes & $\checkmark$ &  OP minimization & Analytical    \\ \hline
			\cite{subhash2022optimal} & SIMO   &  \texttimes & $\checkmark$ & Minimum SINR maximization  & projected gradient descent \\ \hline
			\cite{subhash2022max} & SIMO & $\checkmark$  & \texttimes  &  Minimum SINR maximization & Geometric programming \& matrix-lifting\\ \hline
			\cite{dai2021statistical,zhi2021statistical} &  MIMO & \texttimes  & $\checkmark$ & Ergodic rate maximization  & Genetic Algorithm\\ \hline    
			\cite{han2019large} &  MISO &  $\checkmark$ &  $\checkmark$ &  Ergodic rate maximization & Analytical \\ \hline
			This work &  SIMO  & $\checkmark$ &  $\checkmark$  &  OP minimization \& ergodic rate maximization  &  PSO and MPSO \\ \hline
		\end{tabular}
		\caption {Compare and contrast with closely related existing work}
		\label{Tab: MISO_IRS_lit_summary}
	\end{table}	
	Table \ref{Tab: MISO_IRS_lit_summary} provides a summary of compare and contrast of this work with closely related existing work.
	\subsubsection*{Organization}
	The rest of the paper is organized as follows. Section \ref{Sec: MISO_IRS_sys_model} describes the system model and derives an approximate OP expression. In Section \ref{Sec: MISO_IRS_Optimization_Problem}, we formulate the optimization problem and propose a solution. In Section \ref{Sec: MISO_IRS_Simualtion_Results}, we provide extensive numerical results and study the impact of different parameters on the OP. Finally, Section \ref{Sec: MISO_IRS_conclusion} concludes the work.
	\subsubsection*{Notation} In this paper, $ \operatorname{G}\left(a,b\right) $ denotes the Gamma distribution with shape parameter $a$ and scale parameter $b$. $\mathcal{CN}\left(\mu,\sigma^{2}\right)$ denotes the complex Gaussian distribution with mean $\mu$ and variance $\sigma^{2}$. The mean of a random variable $ X $ is denoted by $\mathbb{E}\left[X \right]$. For a vector $\mathbf{z}$, $\left[ \mathbf{z} \right]_{i}$ denotes its $i$th element and $\left[ \mathbf{X}\right]_{i,j}$ denotes the $i,j$th entry of matrix $\mathbf{X}$. $\operatorname{diag}(a_{1},\cdots,a_{N})$ denotes a diagonal matrix with entries $a_{1},\cdots,a_{N}$.  
	
	\section{System Model} \label{Sec: MISO_IRS_sys_model}
	We consider the uplink of a system consisting of a multiple-antenna base station (BS) having $M$ antennas communicating with a single-antenna user (D) using an IRS with $N\in \mathbb{N}$ reflecting elements. 
	Let,
	$\mathbf{H}^{SR} \in  \mathbb{C}^{M\times N}$, $\mathbf{h}^{RD} \in  \mathbb{C}^{N\times 1}$ and $\mathbf{h}^{SD} \in  \mathbb{C}^{M\times 1}$ denote the small-scale fading channel coefficients of the BS to IRS, IRS to D and BS to D link respectively. It is assumed that all the channels experience independent Rician fading. Hence, we have
	\begin{equation}\label{Eq: MISO_IRS_Channelmodel}
		\begin{aligned}
			&\left[ \mathbf{h}^{SD}\right]_{i} \sim \mathcal{CN}\left(\mu_{sd},\sigma_{sd}^{2} \right)  \\
			&\left[\mathbf{H}^{SR}\right]_{i,j} \sim \mathcal{CN}\left(\mu_{sr}, \sigma_{sr}^{2}\right) \ \forall  i \in \lbrace 1,\cdots,M \rbrace,\forall  j \in \lbrace 1,\cdots,N \rbrace   \\
			&\left[\mathbf{h}^{RD}\right]_{i} \sim \mathcal{CN}\left(\mu_{rd}, \sigma_{rd}^{2}\right) \ \forall  i \in \lbrace 1,\cdots,N \rbrace
		\end{aligned}
	\end{equation}
	where $ \sigma_{ab}^{2} = \frac{d_{ab}^{-\beta_{ab}}}{K_{ab} + 1}, \mu_{ab} = d_{ab}^{-\beta_{ab}/2}\sqrt{\frac{K_{ab}}{K_{ab} + 1}} $ and $ a,b \in \curleybracket*{S,D,R}$\textbf{\add{,}} $\beta_{sd}, \beta_{sr}$ and $ \beta_{rd} $ are the path loss coefficients and $ K_{sd}, K_{sr} $ and $ K_{rd} $ are the Rice factors of respective links. Let $\alpha$ and $\theta_n$ represent the amplitude and phase introduced by the $n$-th IRS element, respectively. Let $\mathbf{f}$ be the combining beamforming vector used by the BS, then the received signal at the BS is given by  
	\begin{equation}
		y = \sqrt{p}\mathbf{f}^{H}\left(\mathbf{h}^{SD}+\mathbf{H}^{SR}\boldsymbol\Theta\mathbf{h}^{RD}\right)s+w,
		\label{Eq: MISO_IRS_sig_recvd}
	\end{equation}
	where $\boldsymbol\Theta = \operatorname{diag} ( \alpha \operatorname{e}^{j\theta_1},....., \alpha \operatorname{e}^{j\theta_N})$, $p$ is transmit power , $s$ is the transmitted signal with $\mathbb{E}[|s|^{2} ]$=1 and $w$ is the AWGN with noise power $\sigma^{2}$. Similar to the authors of \cite{zhi2020uplink}, we used maximum ratio combining  ({MRC}), which is the optimal combining beamforming solution that maximizes the received signal power at the BS for a constant phase shift at the IRS. Hence, we have
	\begin{equation}
		\mathbf{f}=\frac{\mathbf{h}^{SD}+\mathbf{H}^{SR}\boldsymbol\Theta\mathbf{h}^{RD}}{\norm*{ \mathbf{h}^{SD}+\mathbf{H}^{SR}\boldsymbol\Theta\mathbf{h}^{RD}}}
		\label{Eq: MISO_IRS_transmit_beamform}
	\end{equation}
	The SNR for the IRS-assisted system at the BS is then given by
	\begin{equation}\label{Eq: MISO_IRS_snr_1}
		\gamma_{IRS} =  \gamma_s {  \norm*{\mathbf{h}^{SD}+\mathbf{H}^{SR}\Theta\mathbf{h}^{RD}} }^2,
	\end{equation}
	Next, we use the method of moment matching and propose an approximate expression for the PDF of the SNR in \eqref{Eq: MISO_IRS_snr_1}. This result is presented in the following theorem:
	\begin{lemma} \label{Lemma: MISO_IRS_gamma_approx}
		The PDF of $\gamma_{IRS}$ is approximated as 
		\begin{equation}
			\begin{aligned}
				f_{\gamma_{IRS}}\left( x\right) &= \frac{x^{k_{mom}-1} \operatorname{e}^{-x/\theta_{mom}} }{\theta_{mom}^{k_{mom}} \Gamma\left(k_{mom} \right) } 
			\end{aligned}
		\end{equation}
		where the shape parameter $\left( k_{mom} \right)$ and the scale parameter $ \left( \theta_{mom} \right) $ of the Gamma distribution can be evaluated using:
		\begin{equation}\label{Eq: MISO_IRS_Par_k}
			k_{mom} =\frac{\left(\mathbb{E}[\gamma_{IRS}]\right)^{2} }{\mathbb{E}[\gamma_{IRS}^2] -                     \left(\mathbb{E}[\gamma_{IRS}]\right)^{2} },
		\end{equation}
		%%%%%%%%%%%
		\begin{equation}\label{Eq: MISO_IRS_Par_theta}
			\theta_{mom} = \frac{\mathbb{E}[\gamma_{IRS}^2] -                     \mathbb{E}^2[\gamma_{IRS}]}{E[\gamma_{IRS}]}.
		\end{equation}
		Here, $\mathbb{E}[\gamma_{IRS}]$, $\mathbb{E}[\gamma_{IRS}^2]$ are evaluated using (\ref{Eq: MISO_IRS_snr_mean_1}) and  (\ref{Eq: MISO_IRS_SecondMoment_Step_1}) respectively. (see Appendix \ref{App: MISO_IRS_proof_gamma_moment})
	\end{lemma}
	\begin{proof}
		Please refer to Appendix \ref{App: MISO_IRS_proof_gamma_moment} for the proof.
	\end{proof}
	Lemma \ref{Lemma: MISO_IRS_gamma_approx} characterize the PDF of the $ \gamma_{IRS} $ using which other performance metrics of interest, such as OP, ergodic rate, etc., can also be derived. 
	\begin{theorem}\label{Thm: MISO_IRS_OP} 
		The OP of the considered system, for threshold $ \gamma_{th} $, is given as
		\begin{equation}\label{Eq: MISO_IRS_p_out_mom}
			\begin{aligned}
				P_{outage}(\gamma_{th}) = \frac{\gamma\left(k_{mom}, \frac{\gamma_{th}}{\theta_{mom} } \right) }{\Gamma\left( k_{mom} \right)},
			\end{aligned}
		\end{equation}
		Here, $\gamma(\cdot,\cdot)$ is the lower incomplete Gamma function \cite{abramowitz1965:handbook}.
	\end{theorem}
	\begin{proof}
		The OP is defined as 
		\begin{equation}
			\begin{aligned}
				P_{outage}(\gamma_{th}) = \mathbb{P}\left[ \gamma_{IRS} \le \gamma_{th} \right]
			\end{aligned}
		\end{equation}
		From Lemma \ref{Lemma: MISO_IRS_gamma_approx} we have  $ \gamma_{IRS} \sim \operatorname{G}\left(k_{mom} ,\theta_{mom} \right) $, hence
		\begin{equation}
			\begin{aligned}
				P_{outage}(\gamma_{th}) = \frac{1}{\theta_{mom}^{k_{mom} } \Gamma\left(k_{mom} \right) } \int_{0}^{\gamma_{th} } x^{k_{mom} -1} \operatorname{e}^{-x/\theta_{mom} }  dx
			\end{aligned}
		\end{equation}
		Next, using the definition of lower incomplete Gamma function \cite[Eq. 6.5.2]{abramowitz1965:handbook} gives the result in \eqref{Eq: MISO_IRS_p_out_mom}. This completes the proof.
	\end{proof}
	\begin{theorem}\label{Thm: MISO_IRS_Rate}
		The ergodic rate of the considered system is 
		\begin{equation}\label{Eq:MISO_IRS_RateFinal}
			\begin{aligned}
				C &= \frac{1}{\ln\left( 2\right) \Gamma\left(k_{mom} \right) } G_{3,2}^{1,3}\left( \theta_{mom} \ \begin{array}{|c}
					1-k_{mom} ,1,1 \\
					1,0 
				\end{array}  \right)
			\end{aligned}
		\end{equation},
	\end{theorem}
	where $ G^{m,n}_{p,q} \left(x \ \begin{array}{|c}
		a_{1},\cdots,a_{p}  \\
		b_{1},\cdots,b_{q}
	\end{array} \right) $ is Meijer's G function \cite{MeijerGdef}.
	\begin{proof}
		By the definition of ergodic rate, we have
		\begin{equation}
			\begin{aligned}
				C &= \mathbb{E}\left[\log_{2}\left(1 + \gamma_{IRS} \right) \right] 
			\end{aligned}
		\end{equation}
		From Lemma \ref{Lemma: MISO_IRS_gamma_approx} we have  $ \gamma_{IRS} \sim \operatorname{G}\left(k_{mom},\theta_{mom} \right) $, hence
		\begin{equation}\label{Eq:MISO_IRS_Rate1}
			\begin{aligned}
				C &= \frac{1}{\ln\left( 2\right) \theta_{mom}^{k_{mom} } \Gamma\left(k_{mom} \right) }\int_{0}^{\infty} \ln\left(1 + x \right) x^{k_{mom} -1} \operatorname{e}^{-x/\theta_{mom} }  dx
			\end{aligned}
		\end{equation}
		using \cite{weisstein2003logmeigerG} and \cite[Eq. 7.813.1]{Grad2007} , we have 
		\begin{equation}\label{Eq:MISO_IRS_Rate2}
			\begin{aligned}
				\int_{0}^{\infty} \ln\left(1 + x \right) x^{k_{mom} -1} \operatorname{e}^{-x/\theta_{mom} }  dx &= \int_{0}^{\infty} G_{2,2}^{1,2}\left(x \ \begin{array}{|c}
					1,1 \\
					1,0 
				\end{array}  \right) x^{k_{mom} -1} \operatorname{e}^{-x/\theta_{mom} }  dx \\
				&= \theta_{mom}^{k_{mom} } G_{3,2}^{1,3}\left(\theta_{mom} \ \begin{array}{|c}
					1-k_{mom},1,1 \\
					1,0 
				\end{array}  \right)
			\end{aligned}
		\end{equation}
		Substitution of \eqref{Eq:MISO_IRS_Rate2} in \eqref{Eq:MISO_IRS_Rate1} gives the \eqref{Eq:MISO_IRS_RateFinal}, and that completes the proof.
	\end{proof}
	Note that all the above expressions are in closed form\footnote{These expressions can be evaluated easily using in-built functions available in Mathematica and/or Matlab}. Considering the large number of elements present in an IRS, it is suggested to have only finite/discrete phase shift levels since this allows us to use a small number of bits for phase representation, hence making IRS cost-effective \cite{wu2019towards,han2019large,wu2020beamforming}. When one uses $b$ bits to represent the quantized phase, the set of possible phase shifts for each of the $N$ elements is given by 
	\begin{equation}\label{Eq: MISO_IRS_PhaseSet}
		\begin{aligned}
			\mathcal{F} = \curleybracket*{ 0,\frac{2\pi}{2^b},\cdots,\frac{\left(2^b - 1\right)2\pi}{2^b}}. 
		\end{aligned}
	\end{equation}
	In the next section, we utilize these expressions to formulate optimization problems to minimize OP \textit{i.e.,} $P_{outage}$ and maximize the ergodic rate \textit{i.e.,} $C$.
	\section{Optimization problems \& proposed solution} \label{Sec: MISO_IRS_Optimization_Problem}
	In this section, we first formulate an optimization problem to minimize the OP at the destination $\mathbf{D}$. From (\ref{Eq: MISO_IRS_p_out_mom}), we can observe that the OP is a function of phase shifts at IRS \textit{i.e.,} $ \theta_{1},\dots,\theta_{N} $ for any given system model and threshold $ \gamma_{th} $. Hence, our objective is to choose the phase shifts from the available set of discrete phase shifts such that the OP is minimized. Let $\boldsymbol{\theta} = \left[\theta_{1},\dots,\theta_{N} \right]$ denotes the vector containing phase shift at IRS. Mathematically, the formulated optimization problem is given by
	\begin{equation}\label{Eq: MISO_IRS_OptProb}
		\begin{aligned}
			\min_{\boldsymbol{\theta}} \quad &   P_{outage} \\
			\textrm{s.t.} \quad & \theta_n \in \mathcal{F} \ \forall \ n = 1,\dots,N.
		\end{aligned}
	\end{equation}
	where $\mathcal{F}$ is given by \eqref{Eq: MISO_IRS_PhaseSet}. Similarly, from \eqref{Eq:MISO_IRS_RateFinal}, note that the ergodic rate \textit{i.e.,} $C$ is also a function of phase shifts $\boldsymbol{\theta}$ at IRS. Hence, one can aim for a phase shift design that maximizes the ergodic rate $C$. This problem can be mathematically formulated as,
	\begin{equation}\label{Eq: MISO_IRS_EC_OptProb}
		\begin{aligned}
			\max_{\boldsymbol{\theta}} \quad & C  \\
			\textrm{s.t.} \quad & \theta_n \in \mathcal{F} \ \forall \ n = 1,\dots,N.
		\end{aligned}
	\end{equation}
	Since the search space of the optimization problems in \eqref{Eq: MISO_IRS_OptProb} and \eqref{Eq: MISO_IRS_EC_OptProb} is discrete; hence the problem is non-convex. Also, as mentioned earlier, the value of $N$ is large for IRS, so finding the optimal $\boldsymbol{\theta}$ using brute search among $2^{bN}$ feasible solution is not practical. To solve such problems, one can use heuristic algorithms. We propose to use the multi-valued particle swarm optimization (MPSO) algorithm \cite{veeramachaneni2007probabilistically} to solve the \eqref{Eq: MISO_IRS_OptProb} and \eqref{Eq: MISO_IRS_EC_OptProb} since it provides the solution for such problems with time complexity independent of the system parameter. The details of MPSO are given in the Subsection \ref{Subsec: MISO_IRS_MPSOsec}. Further, to investigate the loss in performance due to the unavailability of continuous phase shifts at the IRS, we also solved the optimization problems in (\ref{Eq: MISO_IRS_OptProb}) and \eqref{Eq: MISO_IRS_EC_OptProb} for the special case, when the IRS can adjust to any phase between $ 0 $ to $ 2\pi $ using particle swarm optimization (PSO) algorithm \cite{kennedy1995particle,shi1998modified,kuila2014energy},  details of which are given in Subsection \ref{Subsec: MISO_IRS_PSOsec}.  
	\subsection{Multi-Valued PSO}\label{Subsec: MISO_IRS_MPSOsec}
	The MPSO algorithm \cite{veeramachaneni2007probabilistically} is a modified version of binary PSO (BPSO) \cite{kennedy1997discrete}, which can solve a multi-valued optimization problem without transforming it into an equivalent  binary representation. Similar to any swarm optimization technique, the MPSO algorithm also starts with an initial set of particles sampled from the solution space. This initial set is generally known as the initial ``population'', and each set element is called a ``particle''. In the context of our optimization problems \textit{i.e.,} \eqref{Eq: MISO_IRS_OptProb} and \eqref{Eq: MISO_IRS_EC_OptProb}, a particle is nothing but a particular value of phase shifts at IRS \textit{i.e.,} $\boldsymbol{\theta} = \left[\theta_{1},\dots,\theta_{N} \right]$. Let $T$ particles be used in implementation, then the population at $i$th iteration is denoted as $\boldsymbol{\theta^{\left(i,j \right)}} \ \forall j = 1,\dots, T$.  In every iteration, the objective function (OP or ergodic rate, depending on the problem) is evaluated for each particle. Based on the value of the objective function, the ``position'' or the value of each particle \textit{i.e.,} $\boldsymbol{\theta}^{\left(i,j \right)}$ is updated as follows:
	\begin{equation}\label{Eq: MISO_IRS_VelocityMPSO}
		\mathbf{V}_{i,j} = \Omega \mathbf{V}_{i-1,j} + \psi_{1}\left(\boldsymbol{\theta}^{j}_{local} - \boldsymbol{\theta}^{\left(i,j\right)} \right) + \psi_{2}\left(\boldsymbol{\theta}_{global} - \boldsymbol{\theta}^{\left(i,j\right)} \right), 
	\end{equation}
	where $ \mathbf{V}_{i,j} $ is the velocity (rate of position change) of $j$th particle in $i$th iteration and $ \Omega = 0.9 - \frac{i\left(0.9 - 0.2\right)}{I_{max}} $ is the inertia weight and $ \psi_{1}, \psi_{2} \in \left[0,2\right] $ are two random positive numbers. In \eqref{Eq: MISO_IRS_VelocityMPSO} $\boldsymbol{\theta}^{j}_{local}$ is the best position of $j$th particle and $\boldsymbol{\theta}_{global}$ is the best position of among all the particles, till $i$th iteration. Here, the best position refers to the value of phase shifts at IRS that has minimum OP or maximum ergodic rate, depending on the problem. Hence, the velocity is updated based on the particle's own best experience as well as the population's global best experience. Velocity \textit{i.e.},$ \mathbf{V}_{i,j} $ controls the change in phase shifts of the updated solution. Next, the updated velocity is mapped to the range of solution space using a sigmoid function as follows
	\begin{equation}\label{Eq: MISO_IRS_SigmoidMPSO}
		\begin{aligned}
			\mathbf{S}_{i,j}  &= \dfrac{2^{b} - 1}{1 + \exp\left({-\mathbf{V}_{i,j}}\right) }
		\end{aligned}
	\end{equation}
	The position of each particle is updated using $ \mathbf{S}_{i,j} $ as follows
	\begin{equation}\label{Eq: MISO_IRS_PositionUpdateMPSO}
		\begin{aligned}
			\boldsymbol{\Theta}^{\left(i+1,j\right)} &= \left[\operatorname{round}\left(\mathcal{N}\left(\mathbf{S}_{i,j},\sigma\left(2^{b}-1\right)\right)\right)\right] \frac{2\pi}{2^b}
		\end{aligned}
	\end{equation} 
	After that, the following projection operation is executed to ensure that the updated particle falls in the solution range. 
	\begin{equation}\label{Eq: MISO_IRS_ProjectionMPSO}
		\begin{aligned}
			\theta_{n}^{i+1,j} &= \begin{cases*}
				\frac{\left(2^{b} - 1\right)2\pi}{2^{b}}  \quad \theta_{n}^{i+1,j} > 2^{b}-1\\
				0 \qquad \quad  \ \ \theta_{n}^{i+1,j} < 0 \\
				\theta_{n}^{i+1,j} \qquad  \text{otherwise}\\
			\end{cases*}
		\end{aligned}
	\end{equation}  
	This process is repeated till the stopping criterion is met. The complete algorithm for OP minimization is presented in the Algorithm \ref{Algo: MISO_IRS_MPSO}.
	\begin{algorithm}
		\caption{MPSO Algorithm}
		\begin{algorithmic}[1]\label{Algo: MISO_IRS_MPSO}
			\STATE \textit{\textbf{Initialization}}: Generate $ T $ particles $ \boldsymbol{\theta}^{\left(1,j\right)}, j = 1,\dots,T $
			\FOR{$ i = 1:I_{max}$} 
			\FOR{$ j = 1:T$}
			\STATE Compute the OP using \eqref{Eq: MISO_IRS_p_out_mom} for each $ \boldsymbol{\theta}^{\left(i,j\right)} $ denoted as $ P_{i,j} $
			\ENDFOR
			\STATE Find $ \left(i_{m},j_{m}\right) = \arg \min\limits_{i,j} \ P_{i,j} $. Set $ P_{min}^{i} = P_{i_{m},j_{m}} $ and $ \boldsymbol{\theta}_{global} = \boldsymbol{\theta}^{(i_{m},j_{m})}$ \COMMENT{Finding global best till current iteration}
			\FOR{$ j = 1:T$}
			\STATE  Get $ i_{j} = \arg\min\limits_{i} \ P_{i,j} $ and set $ P_{local}^{j}  = P_{i_{j},j}\ \forall j $ and $ \boldsymbol{\theta}^{j}_{local}  = \boldsymbol{\theta}^{i_{j},j} \ \forall j$ \COMMENT{Finding personal best for each particle till current iteration}
			\STATE Calculate velocity for each particle using \eqref{Eq: MISO_IRS_VelocityMPSO} and
			map it to solution space using \eqref{Eq: MISO_IRS_SigmoidMPSO}
			\STATE Update the particle's position using \eqref{Eq: MISO_IRS_PositionUpdateMPSO}
			\STATE For each $ \theta_{n}^{i+1,j} $ check if $ \theta_{n}^{i+1,j} < 0 $ then $ \theta_{n}^{i+1,j} = 0  $, else if $ \theta_{n}^{i+1,j} > 2\pi  $ then  $ \theta_{n}^{i+1,j} = 2\pi  $
			\ENDFOR
			\ENDFOR
		\end{algorithmic}
	\end{algorithm}
	The complexity of Algorithm \ref{Algo: MISO_IRS_MPSO} is $I_{max}T$ as it requires computing the objective function this many times.  In simulations, we considered $I_{max} = 100$ and $T = 200$, which is independent of other system parameters.  
	\subsection{PSO}\label{Subsec: MISO_IRS_PSOsec}
	The PSO \cite{kennedy1995particle} was first introduced by James Kennedy and Russell Eberhart in $ 1995 $ where a paradigm based on the social behavior model is used to solve non-linear optimization problems. Later a slightly modified version of PSO was presented in \cite{shi1998modified} where the concept of \textit{inertia weight} $ \omega $ was incorporated in velocity update. The authors discussed the impact of $ \omega $ on the performance of PSO through simulation and suggested a preferable range of values for $ \omega $. PSO is also an evolutionary process that starts with an initial set of possible solutions sampled from a feasible set. General heuristics and steps of PSO are similar to MPSO counterparts except for a few modifications. Similar to MPSO, the OP is calculated for the initial population, and then the global and local best are selected. The velocity and position  of each particle are updated as follows:
	\begin{equation}\label{Eq: MISO_IRS_VelocityPSO}
		\mathbf{V}_{i,j} = \omega \mathbf{V}_{i-1,j} + c_{1}r_{1}\left(\boldsymbol{\theta}^{j}_{local} - \boldsymbol{\theta}^{\left(i,j\right)} \right) + c_{2}r_{2}\left(\boldsymbol{\theta}_{global} - \boldsymbol{\theta}^{\left(i,j\right)}  \right), 
	\end{equation}
	\begin{equation}\label{Eq: MISO_IRS_PositionUpdatePSO}
		\begin{aligned}
			\boldsymbol{\theta}^{\left(i+1,j\right)} = \boldsymbol{\theta}^{\left(i,j\right)} + \mathbf{V}_{i,j}
		\end{aligned}
	\end{equation}
	where $ \mathbf{V}_{i,j} $ is the velocity (rate of position change) of $j$th particle in $i$th iteration and $ \omega $ is the inertia weight, $c_{1}, c_{2}$ are the acceleration factors, and $ r_{1}, r_{2} \in \left[0,1\right] $ are two random positive numbers. The values of $\omega, c_{1}$ and $c_{2}$ are taken as recommended in \cite{shi1998modified}.  In \eqref{Eq: MISO_IRS_VelocityPSO}, the definition of $\boldsymbol{\theta}^{j}_{local}$ and $\boldsymbol{\theta}^{j}_{global}$ is the same as described in previous subsection. $ \mathbf{V}_{i,j} $ controls the change in phase shifts of the updated solution.
	This process of updating velocity and position is repeated till the stopping criteria are met. 
	Here, detailed steps are provided for the OP minimization problem, but the same algorithms can also be used for ergodic rate maximization by the following modification.
	\begin{itemize}
		\item In the step $4$ of Algorithm \ref{Algo: MISO_IRS_MPSO}, Compute ergodic rate, \textit{i.e.,} $C$ for each particle.  
		\item In the step $6$ and $8$ of Algorithm \ref{Algo: MISO_IRS_MPSO}, global best and local best are the particle that maximizes the objective function 
	\end{itemize}
	\section{Numerical Results \& Discussion} \label{Sec: MISO_IRS_Simualtion_Results}
	In this section, we investigate the performance of the phase shift design approaches presented in the previous section using simulations. For the performance comparison of different phase shift designs, we used the analytical solution based on instantaneous CSI provided in \cite{mishra2019channel}. The reason for this particular choice is two-fold, 1) it provides a near-optimal performance in terms of received SNR maximization, and 2) it is easy  to implement due to its closed-form solution. We consider a simulation setup such that the BS, IRS, and D are placed at $(0,0), (0,10)$, and $(90,0)$, respectively. The value of amplitude coefficient $\alpha$ is taken as $ 1 $ throughout the simulations.  The path loss factors are chosen to be $\beta_{sd} = \beta_{sr} = \beta_{rd} = 4 $, and Rice factors of different links are assumed to be $ K_{sd} = 5, K_{sr} = 10$ and $ K_{rd} = 20 $. The setup chosen here is similar to the one in \cite{charishma2021outage}.
	\begin{figure}[ht]
		\begin{subfigure}{0.48\textwidth}
			\includegraphics[width=\textwidth]{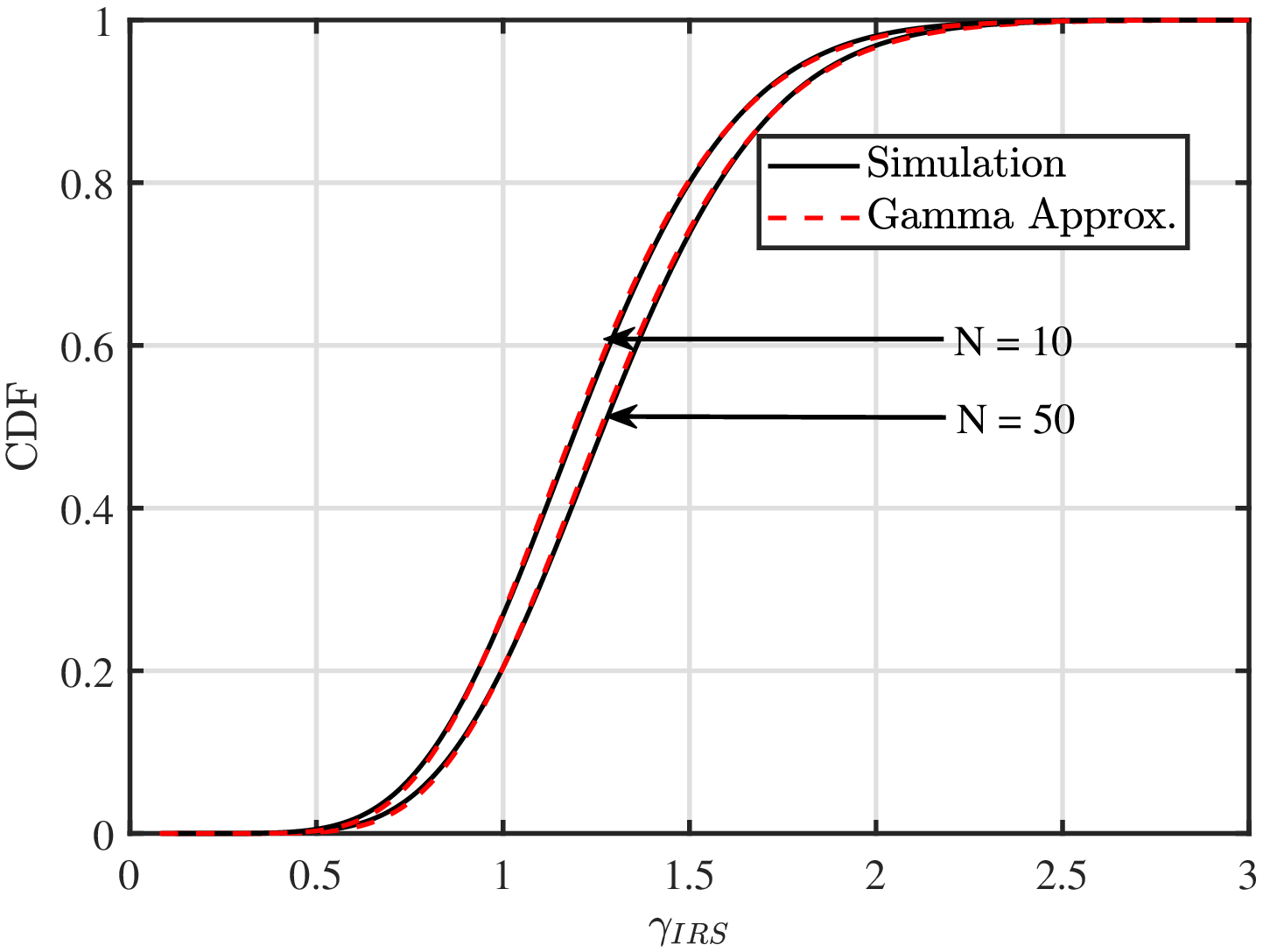}
			\caption{Linear scale}
		\end{subfigure}	
		\hspace{3mm}
		\begin{subfigure}{0.48\textwidth}
			\includegraphics[width=\textwidth]{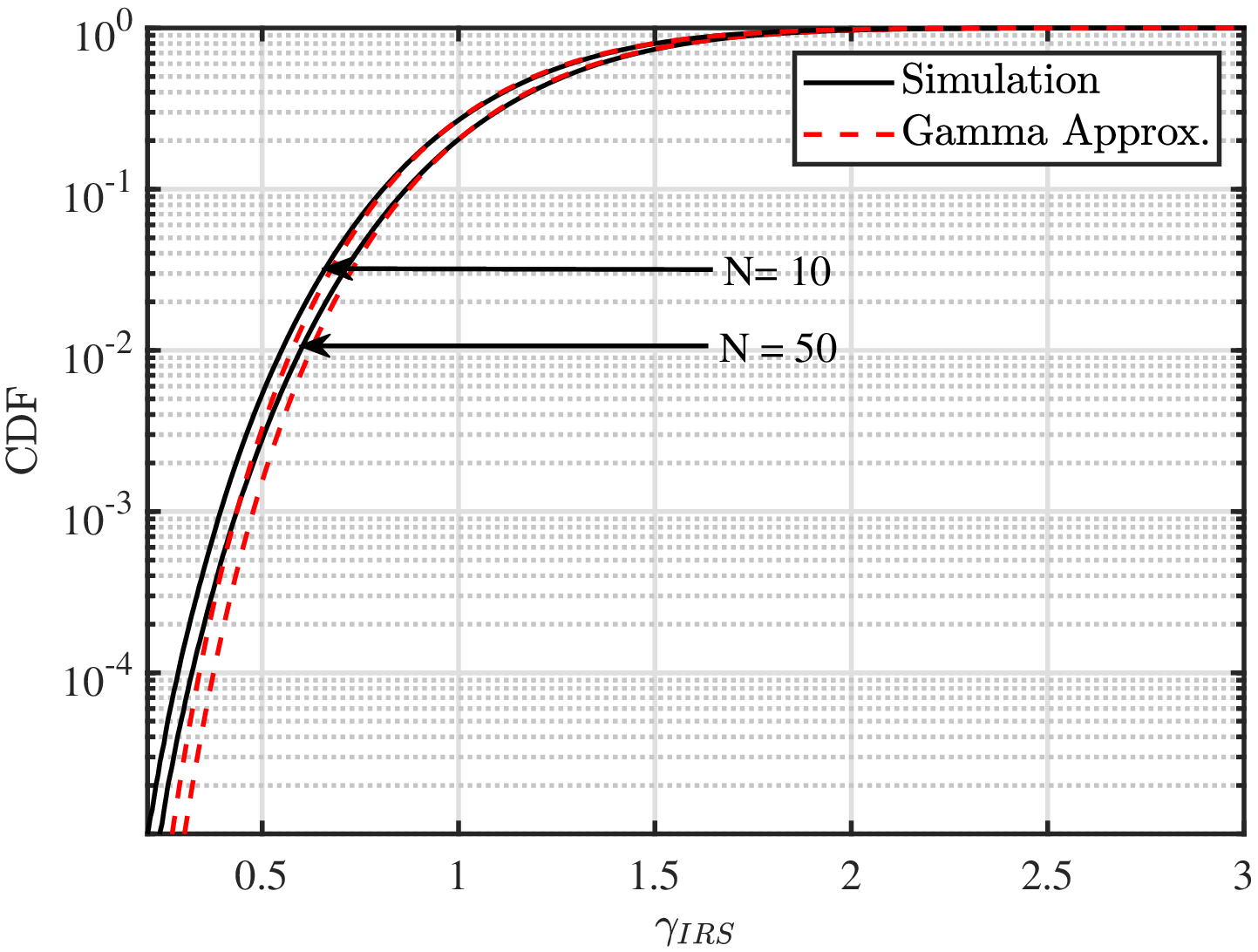}
			\caption{Semi-log scale}
		\end{subfigure}
		\caption{Simulated and approximated CDF of $ \gamma_{IRS} $ for different $N$,  $ d = 0 $ and $ \gamma_{s} = 73 $ dB}
		\label{Fig: MISO_IRS_ApproxVerificationSet1}
	\end{figure}
	Before moving to the comparison of different phase shift designs, we first validate the approximation of OP expression in \eqref{Eq: MISO_IRS_p_out_mom} with simulated values. Fig \ref{Fig: MISO_IRS_ApproxVerificationSet1} (a), (b) plots the CDF of $ \gamma_{IRS} $ for $ d = 0 $ and $ \gamma_{s} = 73 $ dB in linear and semi-log scale, respectively. It is evident from figure \ref{Fig: MISO_IRS_ApproxVerificationSet1} that the Theorem \ref{Thm: MISO_IRS_OP} provides an excellent approximation to the CDF of $ \gamma_{IRS} $.

	\begin{figure}[ht]
		\begin{subfigure}{0.48\textwidth}
			\includegraphics[width=\textwidth]{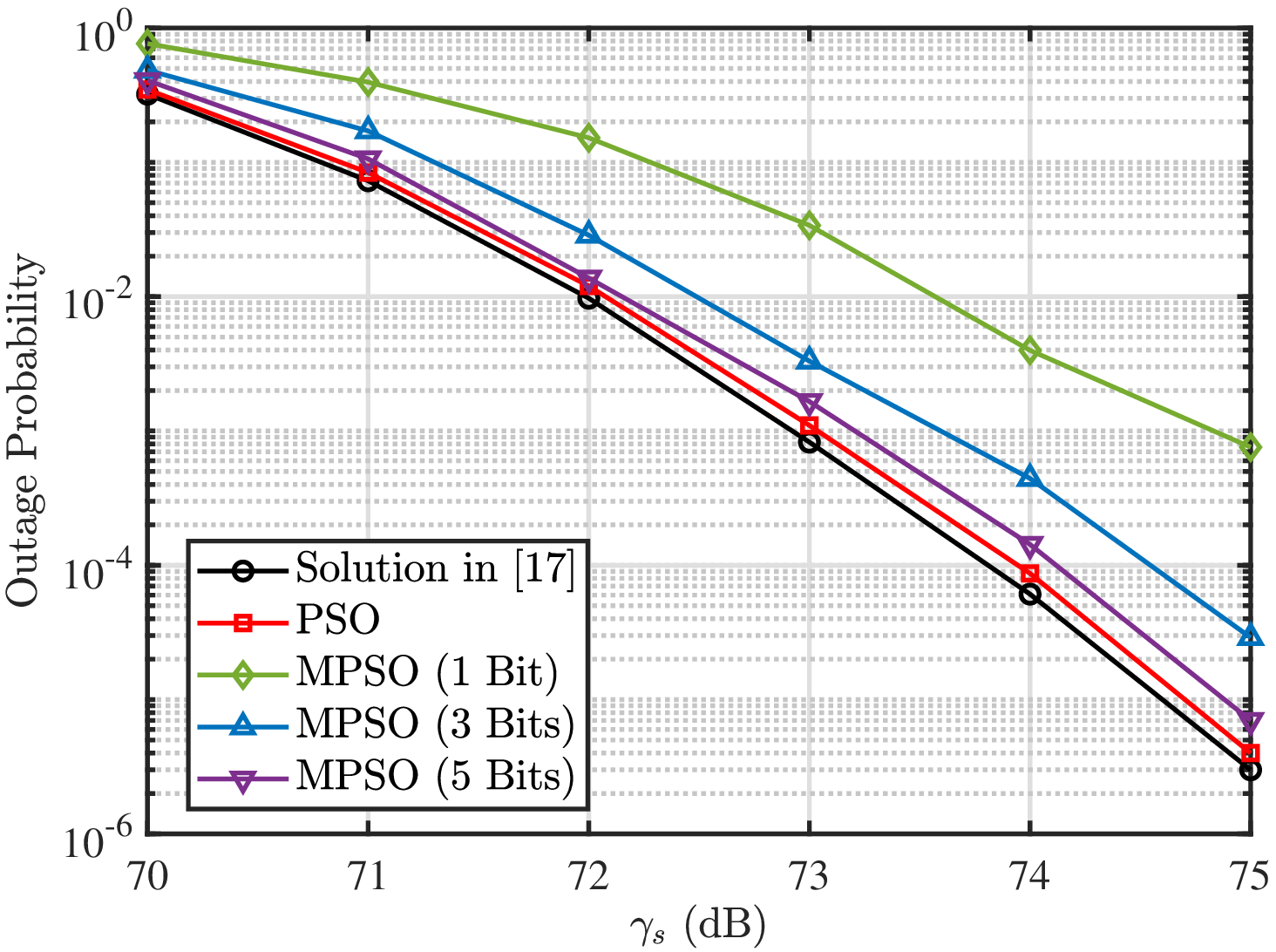}
			\caption{$ N =40 $}
		\end{subfigure}
		\hspace{3mm}
		\begin{subfigure}{0.48\textwidth}
			\includegraphics[width=\textwidth]{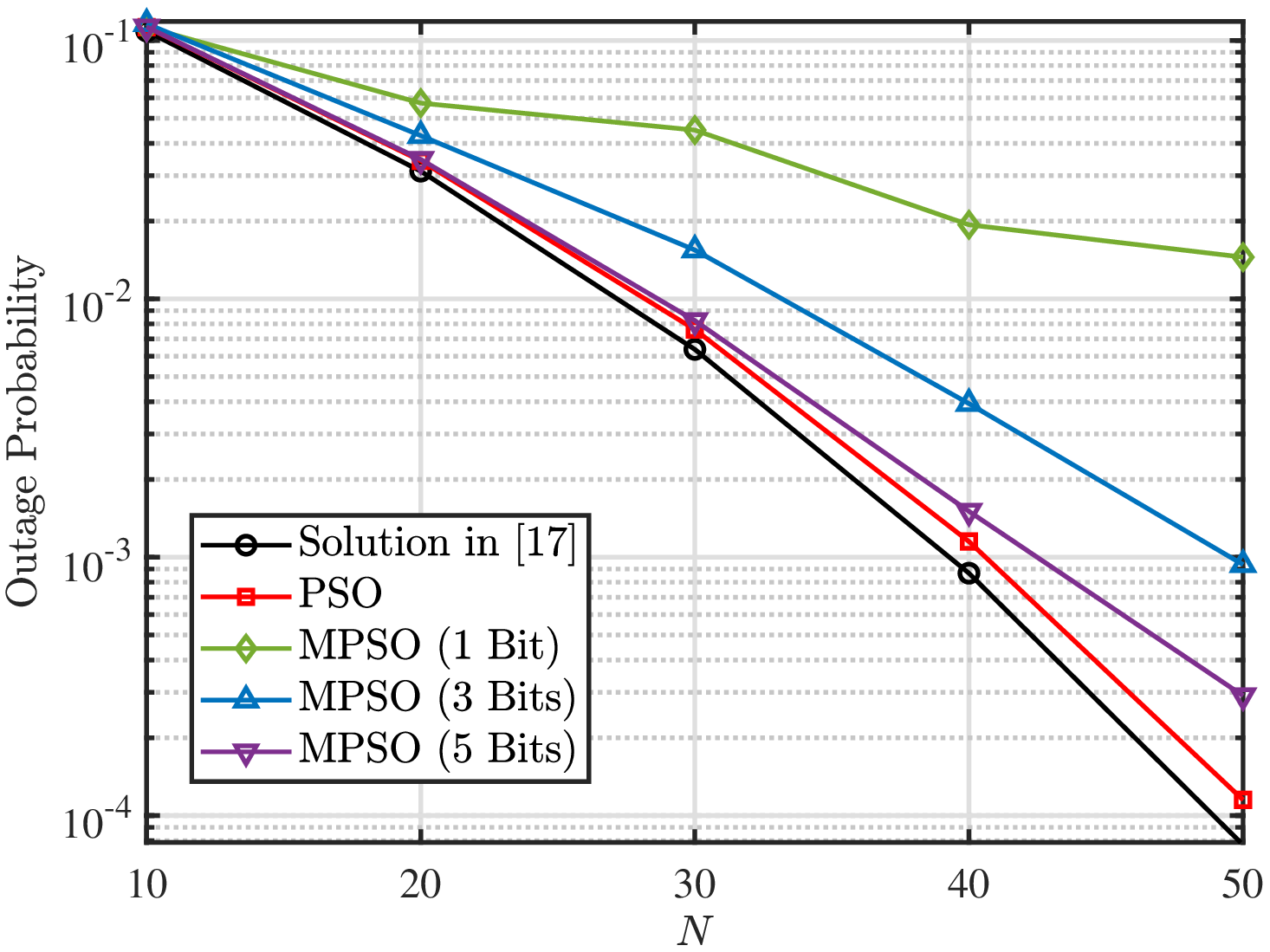}
			\caption{$ \gamma_{s} = 73 $ dB}
		\end{subfigure}
		\caption{Outage Probability versus $\gamma_{s}$ and $N$ for $M = 4$ and $ \gamma_{th} = 0 (dB)$}
		\label{Fig: MISO_IRS_OptCompSetup1}
	\end{figure}
	
	Next, using PSO and MPSO algorithms, we solved the optimization problem in \eqref{Eq: MISO_IRS_OptProb} to obtain the phase shift design at IRS. 
	In Fig. \ref{Fig: MISO_IRS_OptCompSetup1}, we have observed the impact of varying transmit power and number of elements at IRS \textit{i.e.,} $N$ on the OP. It is clear from the numerical results that as the transmit power or $N$ increases, the OP decreases. 
	It is evident from Fig. \ref{Fig: MISO_IRS_OptCompSetup1} (a) and (b) that with an increasing number of bits at IRS, we can achieve better performance. The interesting observation is that the optimized phase shift using PSO and MPSO with $5$ bits can achieve performance close to the one achieved by the solution in \cite{mishra2019channel}. Note that the \cite{mishra2019channel} assumes the knowledge of instantaneous CSI while we use only statistical CSI.
	\begin{figure}[ht]
		\begin{subfigure}{0.48\textwidth}
			\includegraphics[width=\textwidth]{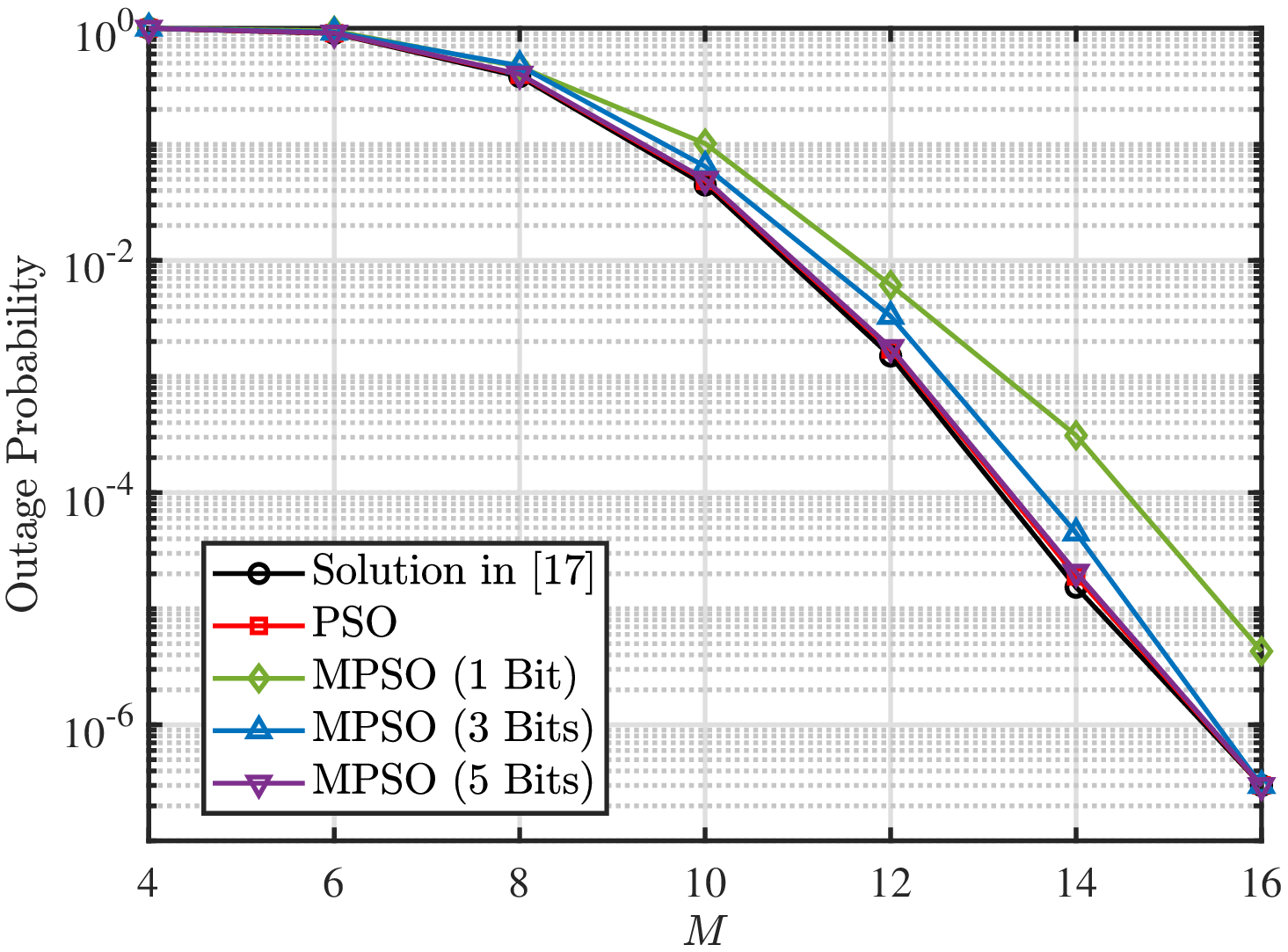}
			\caption{$ \gamma_{th} = 5 $ dB}
		\end{subfigure}
		\hspace{3mm}
		\begin{subfigure}{0.48\textwidth}
			\includegraphics[width=\textwidth]{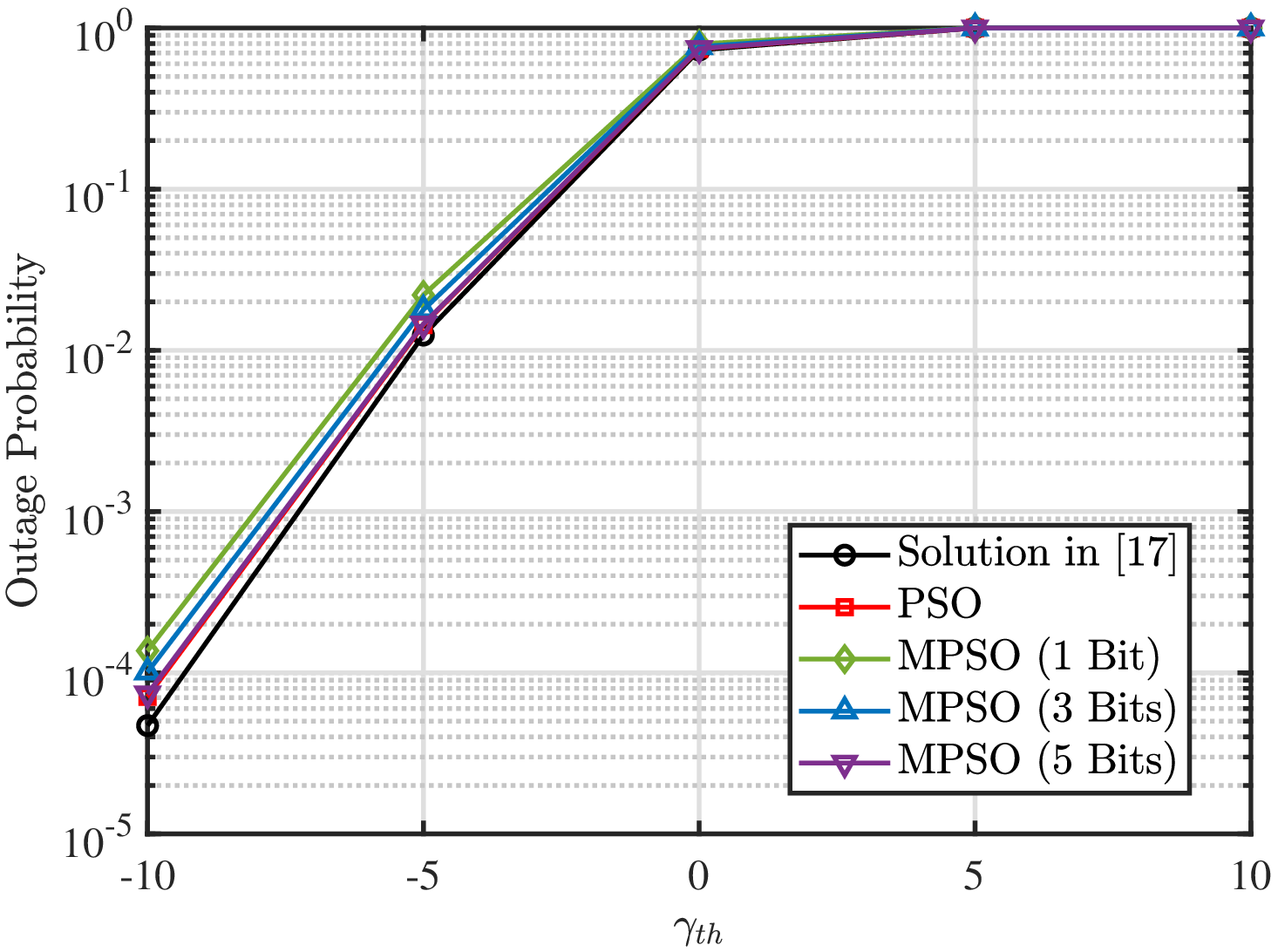}
			\caption{$ M = 2 $ dB}
		\end{subfigure}
		\caption{Outage Probability versus $M$ and $\gamma_{th}$ for $\gamma_{s} = 73 $ dB and $ N = 20 $}
		\label{Fig: MISO_IRS_Opt_Results_M_Th_Vary}
	\end{figure}
	
	Furthermore, we investigated the impact of the number of BS antennae, \textit{i.e.,} $M$ and received SNR threshold, \textit{i.e.,} $\gamma_{th}$ in Fig. \ref{Fig: MISO_IRS_Opt_Results_M_Th_Vary} (a) and (b), respectively with $N = 20$ and $\gamma_{s} = 73$ dB. In Fig. \ref{Fig: MISO_IRS_Opt_Results_M_Th_Vary} (a), as $M$ increases, the OP decreases irrespective of the phase shift design solution due to higher diversity. The OP variation with $\gamma_{th}$ for fixed $M = 2$, $N = 20$ and $\gamma_{s} = 73$ dB is shown in Fig. \ref{Fig: MISO_IRS_Opt_Results_M_Th_Vary} (b). From both the figures, one can observe that the performance obtained through the optimized phase shift using PSO and MPSO with $5$ bits is close to the performance achieved by using \cite{mishra2019channel}. This demonstrates that the statistical CSI-based method can be an alternative to instantaneous CSI-based methods. 
	
	\begin{figure}[ht]
		\begin{subfigure}{0.48\textwidth}
			\includegraphics[width=\textwidth]{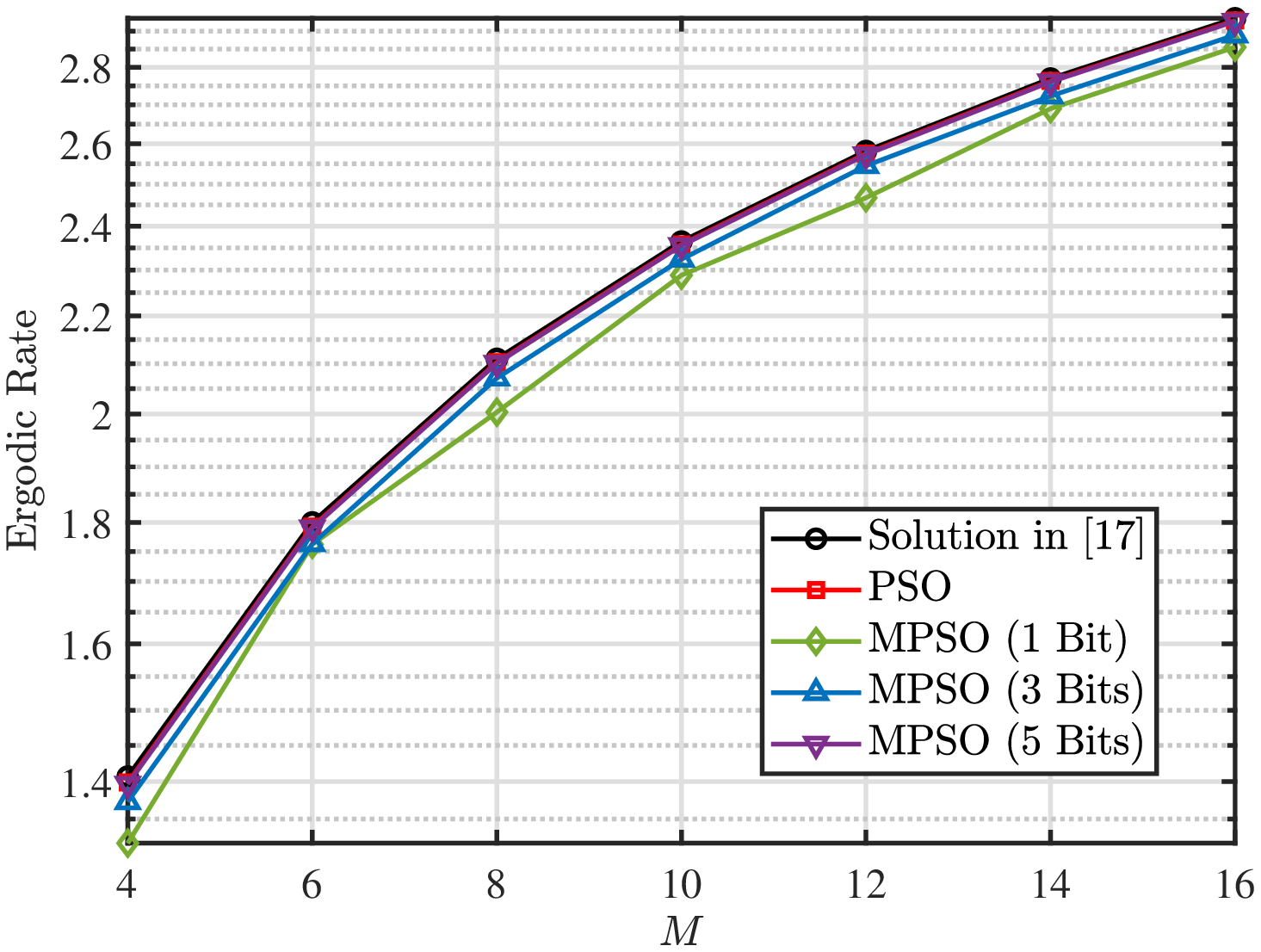}
			\caption{$N = 20, \gamma_{s} = 73 $ dB}
		\end{subfigure}
		\hspace{3mm}
		\begin{subfigure}{0.48\textwidth}
			\includegraphics[width=1.01\textwidth]{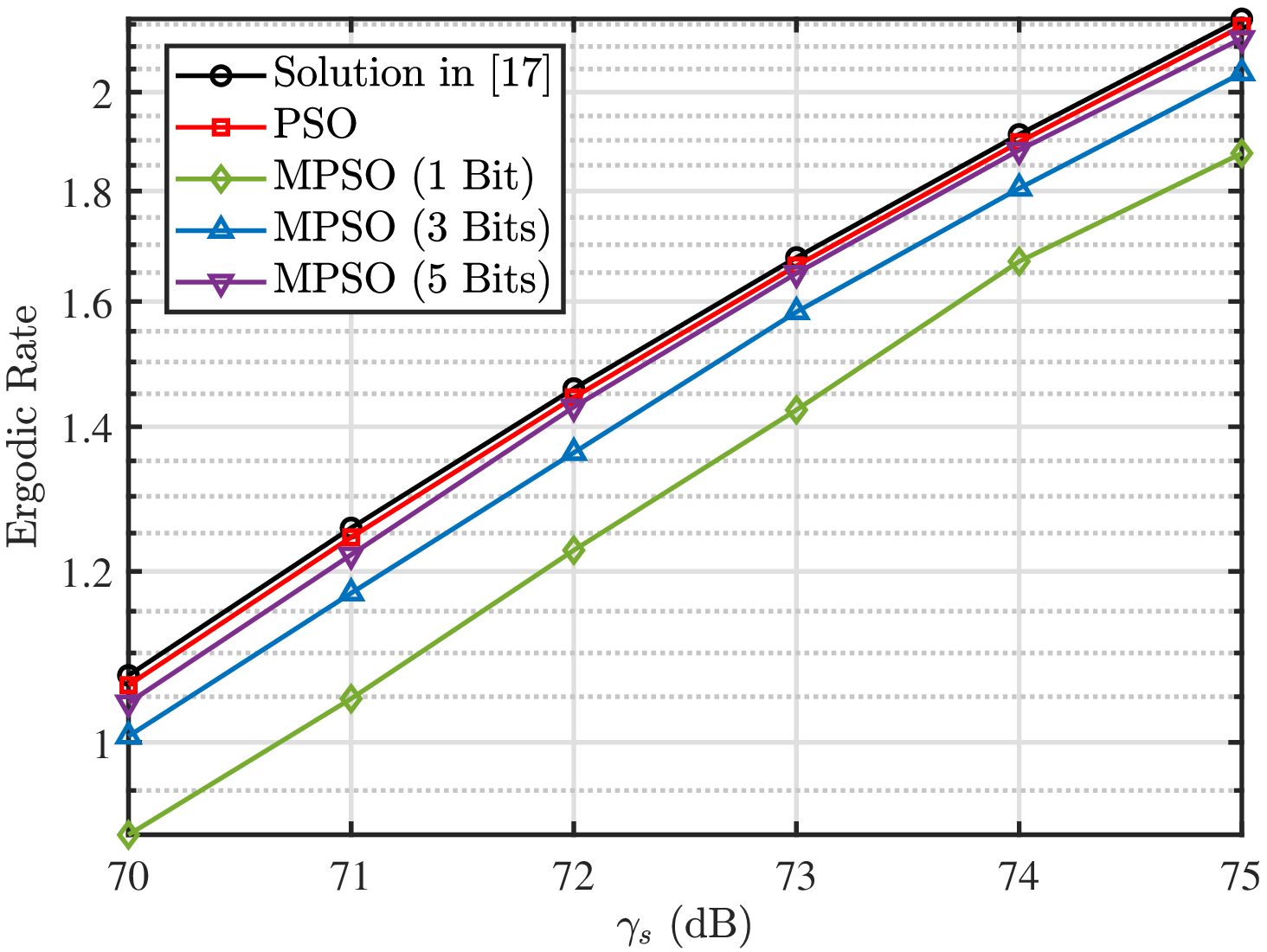}
			\caption{$ M = 4, N = 40 $}
		\end{subfigure}
		\caption{Ergodic rate versus $M$ and $\gamma_{s}$}
		\label{Fig: MISO_IRS_EC_Opt_Results_M_SNR_Vary}
	\end{figure}
	
	In Fig. \ref{Fig: MISO_IRS_EC_Opt_Results_M_SNR_Vary}, we presented the results for ergodic rate with varying $M$ and $\gamma_{s}$. These results also show that maximizing the ergodic rate expression based on statistical CSI, \textit{i.e.,} \eqref{Eq:MISO_IRS_RateFinal} using PSO and MPSO yields as good performance as obtained by near-optimal instantaneous CSI-based phase shift design.
	
	One crucial aspect is that solution in \cite{mishra2019channel} is based on instantaneous CSI, whereas the proposed PSO and MPSO-based solution requires only statistical information about the channel links. The advantage of phase shift design based on statistical CSI is further explained in terms of the reduction of signaling between BS and IRS controller in the subsequent subsection. 
	\subsection{Signalling between BS and IRS controller}\label{Sec: MISO_IRS_Rician_Overhead}
	As mentioned earlier, the IRS elements are programmed using the IRS controller based on the information received from BS. Let's say the continuous phase shifts are represented by $32$ bits; then $32 N$ bits need to be communicated, which can be a considerable overhead as $N$ is typically large. On the other hand, if we use $5$ bits representation for phase shifts, it does not cause any drastic degradation in performance, as shown in various simulation results, then we can reduce the overhead cost. A further reduction in overhead happens because we are only updating phase shifts only when the large-scale fading coefficients are changing. Numerically, if the large-scale fading remains the same for $x$ small-scale fading coherence interval \cite{ashikhmin2018interference,rappaport1996wireless, Ngo2017:CellFree}, then BS needs to update the phase shift only once in lieu of $x$ times. Hence, BS needs to communicate $5N$ bits only compared to $ 32xN $ bits.
	\begin{table}[ht]
		\centering
		% \textcolor{blue}{
			\begin{tabular}{|C{2cm}|C{2cm}|C{2cm}|C{2cm}|C{2cm}|C{2cm}|}
				\hline
				$x$ & $10$ & $20$ & $30$ & $40$  &  $50$   \\ \hline
				Overhead (instantaneous CSI) & $320N$ bits & $640N$ bits & $960N$ bits & $1280N$ bits  &  $1600N$ bits \\ \hline
				Overhead(statistical CSI) & $ 5N $ bits & $ 5N $ bits & $ 5N $ bits & $ 5N $ bits  &  $ 5N $ bits\\ \hline
				Reduction &  $98.44\% $ & $ 99.22\%$ & $99.48\% $ & $99.61\%$  &  $99.69\%$\\ \hline
			\end{tabular} 
			\caption{Overhead reduction for BS to IRS controller signaling}
			\label{Tab: MISO_IRS_Overhead_Reduction} 
			% }
	\end{table}
	Table \ref{Tab: MISO_IRS_Overhead_Reduction} shows the overall reduction in signaling between BS and IRS. Note that even if the large-scale fading coefficient remains constant for as low as $10$ small-scale fading coherence intervals, one can achieve a reduction of $98.44\%$ in signaling between BS and IRS. The overall reduction can go as high as $99.69\%$ if the large-scale fading coefficient remains constant for $50$ small-scale fading coherence intervals.
	\section{Conclusion} \label{Sec: MISO_IRS_conclusion}
	In this work, we have studied the uplink of an IRS-assisted SIMO communication system. We have derived the closed-form expression for outage probability (OP) and ergodic rate based on statistical CSI. The derived expressions are used to design the phase shift for IRS such that the OP is minimized and the ergodic rate is maximized. Our simulation results show that the performance of the statistical CSI-based design closely matches the one with instantaneous CSI. The impact of quantized phase shifts is also studied, and it is shown through extensive simulation that with $5$ bits quantization level, the performance loss is negligible. We also discussed the impact of statistical CSI-based phase shift design and quantized phase shift on the reduction of overhead between BS and IRS controller. It was shown that the overhead could be reduced up to $99.69\%$ if the large-scale fading coefficient remains constant for $50$ small-scale fading coherence intervals without significant loss in performance.   
	\appendices	
	\section{Gamma moment matching} \label{App: MISO_IRS_proof_gamma_moment}
	After some simple algebraic manipulations, the SNR in (\ref{Eq: MISO_IRS_snr_1}) can be re-written as:
	\begin{equation}\label{Eq: MISO_IRS_expanded_eq}
		\begin{aligned}
			\gamma_{IRS} &= \gamma_s\left(A + 2\operatorname{Re} \left(B\right) + C_{1} + C_{2}\right),
		\end{aligned}
	\end{equation}
	where, $ A = \sum\limits_{i=1}^{M} \abs*{h^{SD}_{i}}^2, B =  \sum\limits_{j=1}^{M} \sum\limits_{i=1}^{N} ({h^{SD}_j})^{H} h^{SR}_{ji}h^{RD}_{i} \nu_{i} $, $C_{1} = \sum\limits_{j=1}^{M} \sum\limits_{i=1} ^{N}  \abs*{{h^{SR}_{ji}}}^2 \abs*{{h^{RD}_i}}^2 $ and $ C_{2} = \sum\limits_{j=1} ^{M} \sum\limits_{i=1} ^{N} \sum\limits_{k\ne i}^{N} \left({h^{SR}_{ji}}\right)^{H} \left({h^{RD}_i}\right)^{H} {\nu_i}^{H} h^{SR}_{jk}h^{RD}_k \nu_k $. Next, we calculated the first and second moments of $ \gamma_{IRS} $.
	\subsection{First Moment}
	The first moment of $\gamma_{IRS}$ is
	\begin{equation}\label{Eq: MISO_IRS_FirstMoment_step_1}
		\begin{aligned}
			\mathbb{E}\left[\gamma_{IRS}\right]
			&= \gamma_s\left(\mathbb{E}\left[A\right]+ 2\operatorname{Re} \left(\mathbb{E}\left[B\right]\right)+\mathbb{E}\left[C_{1}\right] + \mathbb{E}\left[C_{2}\right]\right)
		\end{aligned}
	\end{equation}
	Taking the term-by-term expectation and using the fact that the channel coefficient over the different links has independent Rician fading given in \eqref{Eq: MISO_IRS_Channelmodel}, we have
	\begin{equation}\label{Eq: MISO_IRS_snr_mean_1}
		\begin{aligned}
			\mathbb{E}\left[\gamma_{IRS}\right] &= \gamma_{s}M\left(  d_{sd}^{-\beta_{sd}}  +  2 \mu_{sd}\mu_{sr}\mu_{rd}  \operatorname{Re} \left(  s_{1}\right) \right. \\ &\left. +Nd_{sr}^{-\beta_{sr}} d_{rd}^{-\beta_{rd}} + \left(\mu_{sr}\mu_{rd}\right)^{2} s_{2} \right)
		\end{aligned}
	\end{equation}
	where, $ s_{1} = \sum\limits_{i=1} ^{N} \nu_i $ and $ s_{2} = \sum\limits_{i=1}^{N} \sum\limits_{k\ne i}^{N} (\nu_i)^{H} \nu_k$.
	\subsection{Second Moment}
	From (\ref{Eq: MISO_IRS_expanded_eq}), the second moment of $\gamma_{IRS}$ is
	\begin{equation}\label{Eq: MISO_IRS_SecondMoment_Step_1}
		\begin{aligned}
			\mathbb{E}\left[ \gamma_{IRS}^2\right] &= \gamma_{s}^2\Big(\mathbb{E}\left[A^{2}\right] + 2\operatorname{Re}\left(\mathbb{E}\left[B^{2}\right]\right) +  \mathbb{E}\left[C_{1}^{2}\right] +  \mathbb{E}\left[C_{2}^{2}\right]  \\
			&  + 4 \operatorname{Re}\left(\mathbb{E}\left[AB\right]\right)  + 2\mathbb{E}\left[AC_{1}\right] + 2\mathbb{E}\left[AC_{2}\right]   + 2 \mathbb{E}\left[\abs*{B}^{2}\right] \\
			& + 4 \operatorname{Re}\left(\mathbb{E}\left[BC_{1}\right]\right) + 4 \operatorname{Re}\left(\mathbb{E}\left[BC_{2}\right]\right) + 2 \mathbb{E}\left[C_{1}C_{2}\right] \Big). 
		\end{aligned}
	\end{equation}
	The expectation of each term in the above expression is as follows 
	\begin{equation}
		\begin{aligned}
			\mathbb{E}\left[A^2\right] 
			&= M d_{sd}^{-2\beta_{sd}} \left[ \left(\frac{2K_{sd}+1}{\left(K_{sd} + 1\right)^{2}} \right)  + M \right]
		\end{aligned}
	\end{equation}
	\begin{equation}
		\begin{aligned}
			\mathbb{E}\left[B^{2}\right] &= M^{2} \left( \mu_{sd}\mu_{sr}\mu_{rd} s_{1}\right)^{2}
		\end{aligned}
	\end{equation}
	\begin{equation}\label{Eq:MISO_IRS_mean_C1_2}
		\begin{aligned}
			\mathbb{E}\left[C_1^2\right] &= M N d_{sr}^{-2\beta_{sr}}d_{rd}^{-2\beta_{rd}} \left[ \frac{2K_{sr}+1}{\left(K_{sr} + 1\right)^{2}} \left(\frac{2K_{rd}+1}{\left(K_{rd} + 1\right)^{2}} + 1\right)  \right. \\
			&\left. + M\left(\frac{2K_{rd}+1}{\left(K_{rd} + 1\right)^{2}} + N\right) \right]
		\end{aligned}
	\end{equation}
	\begin{equation}\label{Eq:MISO_IRS_mean_C2_2}
		\begin{aligned}
			\mathbb{E}\left[C_{2}^{2} \right] &=  MN\left(N-1\right) \sigma_{sr}^{4} d_{rd}^{-2\beta_{rd}} \left[1  +  \frac{2 K_{sr} }{\left(K_{sr} + 1 \right)^{2}} + \frac{ M K_{sr}^{2} }{\left(K_{rd} + 1\right)^{2}}  \right]       \\
			&+ M\sigma_{sr}^{2} \mu_{sr}^{2} d_{rd}^{-\beta_{rd}} \mu_{rd}^{2} \left(s_{3} + s_{4}\right)\left[1 +  M\frac{K_{sr}}{\left(K_{rd} + 1\right)}\right]\\
			&+ M^{2}\sigma_{sr}^{4}\sigma_{rd}^{4} \left(s_{2}\right)^{2}
		\end{aligned}
	\end{equation}
	where $ s_{3} = \sum\limits_{i=1}^{N} \sum\limits_{k\ne i}^{N} \sum\limits_{w \ne k}^{N}  {\nu_i}^{H}  \nu_w  $ and $ s_{4} = \sum\limits_{i=1} ^{N} \sum\limits_{k \ne i}^{N} \sum\limits_{\substack{v \ne i}}^{N} \nu_k {\nu_v}^{H} $
	\begin{equation}
		\begin{aligned}
			\mathbb{E}\left[AB\right] 
			&= M d_{sd}^{-\beta_{sd}}\mu_{sd}\mu_{sr}\mu_{rd} s_{1}\left[M + \frac{1}{\left(K_{sd} + 1\right)} \right] \\
			&= \mathbb{E}\left[A\right]\mathbb{E}\left[B\right] \left[1 + \frac{1}{M\left(K_{sd} + 1\right)} \right]
		\end{aligned}
	\end{equation}
	Since $A$ and $C_{1}$ are independent hence, we have
	\begin{equation}
		\begin{aligned}
			\mathbb{E}\left[AC_1\right] &= \mathbb{E}\left[A\right]\mathbb{E}\left[C_1\right] =M^{2}N d_{sd}^{-\beta_{sd}} d_{sr}^{-\beta_{sr}} d_{rd}^{-\beta_{rd}}
		\end{aligned}
	\end{equation}
	Again, due to the independence of  $A$ and $C_{2}$, we have
	\begin{equation}
		\begin{aligned}
			\mathbb{E}\left[AC_2\right] &= \mathbb{E}\left[A\right]\mathbb{E}\left[C_2\right] = M^{2} d_{sd}^{-\beta_{sd}} \left(\mu_{sr}\mu_{rd}\right)^{2} s_{2}
		\end{aligned}
	\end{equation}
	After multiplying $B, B^{*}$ and taking the term by term expectation, we have  
	
	\begin{equation}
		\begin{aligned}
			\mathbb{E}\left[\abs*{B}^{2}\right] &= MNd_{rd}^{-\beta_{rd}} \left[ d_{sd}^{-\beta_{sd}} d_{sr}^{-\beta_{sr}}  +\left(M-1\right) \mu_{sd}^{2}\mu_{sr}^{2} \right] \\
			&+ M \mu_{sr}^{2}\mu_{rd}^{2}s_{2} \left[\sigma_{sd}^{2} + M \mu_{sd}^{2} \right]  
		\end{aligned}
	\end{equation}
	\begin{equation}
		\begin{aligned}
			\mathbb{E}\left[BC_{1}\right] &= M  d_{sr}^{-\beta_{sr}} d_{rd}^{-\beta_{rd}}\mu_{sd}\mu_{sr}\mu_{rd}\left[ \frac{1}{K_{sr} + 1}  \right. \\
			&\left. + \frac{1}{K_{sr} + 1} \frac{1}{K_{rd} + 1} +  \frac{M}{K_{rd} + 1} + MN \right] s_{1} 
		\end{aligned}
	\end{equation}
	\begin{equation}
		\begin{aligned}
			\mathbb{E}\left[BC_{2}\right] 
			&= M \mu_{sd}\mu_{sr}\sigma_{sr}^{2}\mu_{rd} d_{rd}^{-\beta_{rd}}\left[  s_{5} + \frac{M K_{sr} }{\left(K_{rd} + 1\right)} s_{5} \right. \\
			& \left.  + \frac{ M K_{sr} K_{rd} }{\left(K_{rd} + 1\right)} s_{1} s_{2} \right] 
		\end{aligned}
	\end{equation}
	
	where $ s_{5} = \sum\limits_{i=1}^{N} \sum\limits_{w\ne i}^{N} \nu_w $
	
	\begin{equation}\label{Eq:MISO_IRS_mean_C1C2}
		\begin{aligned}
			\mathbb{E}\left[C_{1}C_{2}\right] 
			&= Md_{sr}^{-\beta_{sr}} \mu_{sr}^{2} d_{rd}^{-\beta_{rd}} \mu_{rd}^{2} s_{2} \left[ M\left(N+1\right)  + \frac{1}{\left(K_{rd} + 1\right)} \right. \\
			&\left. + \frac{1}{\left(K_{rs} + 1\right)} + \frac{1}{\left(K_{sr} + 1\right) \left(K_{rd} + 1\right)} \right]
		\end{aligned}
	\end{equation}
	\bibliographystyle{IEEEtran}
	\bibliography{refMISO_IRS}
\end{document}